\documentclass[pdflatex,sn-mathphys-num]{sn-jnl}

\usepackage{mathrsfs}
\usepackage{graphicx}%
\usepackage{multirow}%
\usepackage{amsmath,amssymb,amsfonts}%
\usepackage{amsthm}%
\usepackage{mathrsfs}%
\usepackage[title]{appendix}%
\usepackage{xcolor}%
\usepackage{textcomp}%
\usepackage{manyfoot}%
\usepackage{booktabs}%
\usepackage{algorithm}%
\usepackage{algorithmicx}%
\usepackage{algpseudocode}%
\usepackage{listings}%
\usepackage{amsmath}    
\usepackage{braket}     
\usepackage{cleveref}   
\usepackage{natbib}
\usepackage{multirow}   
\usepackage{tabularx}   

\theoremstyle{thmstyleone}%
\theoremstyle{thmstyletwo}%

\theoremstyle{thmstylethree}%

\begin{document}
	
	\title{HTM: A Hybrid Classical-Quantum Computation Model}
	
	
	\author[1]{\fnm{Guoqiang}\sur{Shu}}
	\equalcont{These authors contributed equally to this work.}
	\author[1]{\fnm{Zongjiang} \sur{Yi}}
	\equalcont{These authors contributed equally to this work.}
	\author[1]{\fnm{Jinyang} \sur{Yao}}
	
	\author[1]{\fnm{Haomiao} \sur{Dong}}
    \author[1]{\fnm{Zhihang} \sur{Li}}
	
	\author*[1]{\fnm{Fudong} \sur{Liu}}\email{lwfydy@126.com}
	
	\author*[1]{\fnm{Zheng} \sur{Shan}}\email{shanzhengzz@163.com}
	
	\affil*[1]{\orgname{Information Engineering University}, \orgaddress{\city{Zhengzhou}, \postcode{450001}, \country{China}}}
	
	
	\abstract{Classical and quantum computers have achieved collaborative computation and completed applications across numerous fields. Nevertheless, the absence of a unified hybrid computation model has led to issues such as non-standardized engineering architecture design, ambiguous task partitioning criteria and so on. To address these issues, this research proposes a hybrid classical-quantum computation model (HTM). By constructing unified state sets and transition function representations, HTM clarifies hybrid control logic and implements a control-sharing mechanism. Simultaneously, a comprehensive complexity calculation method based on entanglement and superposition was proposed, supporting computation task partitioning and resource scheduling. The HTM fills a theoretical gap in the classical-quantum hybrid computation area and implements the joint promotion and development of computing models and hardware.}

	\keywords{Hybrid Classical - Quantum computation model, Unified state-transition formalism, Computation complexity, Control-sharing.}
	
	
	
	\maketitle
\section{Introduction}
Currently, the co-integration of classical and quantum computation has become a key enabler for driving quantum computation toward engineering and practical applications. Internationally, representative frameworks including IBM Qiskit, Google Cirq, and NVIDIA, Origin Quantum Sinan, enabling cross-platform reuse of quantum algorithms, supporting algorithm development for NISQ devices, and constructing GPU-QPU collaborative architectures\cite{adarsh2025evolution,caldwell2025platform,fingerhuth2018open}, linking quantum software and hardware. In the context of hybrid computation, global supercomputing centers have integrated quantum and classical computation platforms, establishing collaborative devices. Classical supercomputers manage data preprocessing, error correction, and analysis, while quantum devices excel at solving computationally intractable high-complexity problems such as combinatorial optimization and quantum simulation\cite{sseelam2026reference,rebentrost2024quantum}. These collaborative devices enable intelligent, on-demand scheduling of computation resources and leverages their complementary strengths, in fields such as financial portfolio optimization and biomolecular simulation\cite{pal2024quantum,feynman1982simulating}. Overall, the integrated hardware-software framework of the ``classical + quantum'' has taken initial shape, propelling quantum computation from laboratory theoretical research to practical engineering applications. However, this integrated development still lags in theoretical models behind engineering practice.

The evolution of computational models follows the principle of breaking through the limits of computing capacity imposed by physical devices. Originating in the 1930s, classical computation models focusing on ``computability'', were underpinned by Alan Turing's Turing machine and Alonzo Church's $\lambda$-calculus, which jointly established the Church-Turing thesis and the classical computation framework. Combined with classical devices, the von Neumann architecture laid the foundation for modern classical computers, and device development has driven model optimizations such as parallel Turing machines. Quantum computation models emerged due to the limitations of classical simulations of quantum systems. In the 1980s, Richard Feynman pioneered the concept of employing quantum systems for computation\cite{draeger2017quantum}. In 1985, David Deutsch's quantum Turing machine established quantum computation theoretical framework\cite{deutsch1985quantum}, followed by quantum circuits model adapted to physical implementations (e.g., superconducting circuits, ion traps), leveraging quantum superposition and entanglement to break classical computation efficiency bottlenecks. Computation models and hardware empower each other bidirectionally. However, the absence of a unified hybrid computation model gives rise to a series of critical issues, including non-standardized engineering architectures, inefficient resource coordination, ambiguous task partitioning, elevated cross-platform adaptation costs, and so on. Consequently, task partitioning between classical and quantum resources is currently guided by empirical engineering practices rather than rigorous theoretical principles, which may lead to avoidable resource underutilization and wastage. Thus, a standardized theoretical model supporting the deep integration of quantum computation and supercomputing is imperative.

To address the practical needs of overcoming current technical bottlenecks, this research proposes a hybrid computation model that aims to clarify the core control-sharing mechanism, target the critical issues and establish a complexity analysis method. The hybrid classical-quantum computation model(HTM) defines control-sharing mechanism via unified state set Q and transition function $\delta$\cite{deutsch1985quantum}, clarifying the instruction-dispatching logic from the unified controller to both classical and quantum read-write head. In turn, it provides a standardized theoretical description for HTM, and enables the model to guide engineering practice, resolving inconsistencies in current integration devices. Furthermore, the model adopts a unified complexity description based on quantum state transformation and supports fine-grained, highly reliable scheduling of classical-quantum hybrid resources. This model mitigates bottlenecks in classical computing while guaranteeing efficient, orderly and stable hybrid scheduling for task flows.

In summary, this model fills the theoretical gap in hybrid computation, making a significant contribution to integrated development of quantum and classical supercomputing. Such as it accumulates core technical expertise in control coordination, task partitioning, and interface standardization for universal fault-tolerant quantum computer development, representing an optimal path for deep ``classical + quantum'' integration in the coming period.

\section{Related Work}
The classical Turing machine(TM), proposed by Turing in 1936 as the key classical computation model, laid the theoretical foundation for modern electronic computers. TM defines the boundaries of classical computation and plays an irreplaceable foundational role in constructing the theoretical framework of computer science, as well as in guiding hardware design and algorithm optimization for classical computers. The quantum Turing machine (QTM), as an extension of the classical Turing machine within the framework of quantum mechanics, breaks through the complexity barriers of classical computation. Leveraging quantum properties, QTM offers the potential for exponential acceleration in solving specific problems, becoming the widely accepted model of quantum computation theory. QTM drives the development of quantum algorithms and the engineering exploration of quantum computers.

Meanwhile, complexity theory serves as a key tool for quantifying the computational efficiency of computational models. It clearly delineates efficiency disparities between different models over a range of problems and informs the design and optimization of algorithms. 

\subsection{Classical Computation Model and Complexity}

The classical computation models form the cornerstone of computer science, progressing from mathematical logic to physical machines construction, and ultimately to modern models.

\subsubsection{Classical Computation Model}

Research on formal systems and computability emerged primarily in the 1930s, with four major representative achievements. In 1931, Kurt Gödel proposed Gödel's Incompleteness Theorems, revealing inherent limitations of formal systems and the existence of un-computable problems, while also introducing the recursive function model. In 1936, Alonzo Church presented the $\lambda$-definability model to formalize computability and proved its equivalence with the recursive function model\cite{church1936unsolvable}. The same year, A. M. Turing introduced the Turing machine model, clarifying computability and proving its equivalence to the $\lambda$-definability model. Stephen Kleene proposed recursive function theory—defining recursive functions via elementary function combinations and proving the equivalence of Turing machines, $\lambda$-calculus (later forming the Church–Turing Thesis), thus establishing the boundaries of ``computable functions'' \cite{kleene1981combinatorial}. In the 1930s-1940s, model equivalence and automata theory prevailed. Emil Post's 1936 POST machine (Turing-equivalent) laid the foundation for finite automata theory\cite{post1943formal}. Andrey Markov Jr.'s 1947 Markov algorithm formalized unsolvable problems and advanced language theory\cite{monk1976markov}. From the 1950s to 1970s, abstract models were refined. The Church–Turing Thesis established ``Turing completeness'' and model equivalence\cite{button2009sad,blum2011computer}. Andrey Markov's Markov Chain influenced probabilistic computation theories\cite{lorek2025markov,hoang2023markov}. Calvin C. et al.'s 1964 RASP model formalized practical computer \cite{elgot1964random}. Stephen A. Cook et al.'s 1973 time-bounded RAM model defined inter-model complexity boundaries\cite{cook1973time}.

Among numerous computation models, the Turing machine (TM) forms the cornerstone of theoretical computer science, which defining computability and how computation resources are measured. The Church-Turing Thesis asserts that any ``mechanical'' computation can be simulated by a Turing machine. A TM consists of a finite-state controller, an infinite tape (input/working space), and a read-write head\cite{monk1976markov}. The core research focuses on time complexity (number of execution steps) and space complexity (number of tape cells occupied). The classical Turing machine is the cornerstone of computation theory, and its formal specification is most commonly presented in the form of a 7-tuple (or 5-tuple, depending on the specific contents).

\begin{equation}M = (Q, \Sigma, \Gamma, \delta, q_0, B, F)\end{equation}

Where $\delta$ is Transition Function. Formally, it is a function (or mapping)

\begin{equation}\delta : Q \times \Gamma \mapsto Q \times \Gamma \times \{L, H, R\}\end{equation}

\subsubsection{Classical Complexity}
In classical computation theory, computational complexity is generally quantified in terms of state(or data) transitions\cite{cook1971complexity}. Classical complexity theory takes the consumption of computational resources as its metric, and hierarchically classifies computable problems via the functional relationships among time, space, and input size, forming a theoretical framework for measuring the intrinsic difficulty of problem-solving.

From the perspective of Turing machines, time complexity is defined as the number of steps required to complete computation on an input string, where each step corresponds to a state transition. Thus, time complexity can be interpreted as the path length from the initial state to the halting state in the transition graph\cite{erciyes2021introduction}. Similarly, space complexity refers to the amount of memory used during execution, and is characterized by the number of state records that must be maintained and the number of tape cells accessed throughout the computation.

Under deterministic Turing machines, P consists of polynomial-time solvable decision problems, NP of polynomial-time verifiable ones (SAT and 3-SAT are NP-complete)\cite{adleman1978two}. PSPACE includes polynomial-space decidable problems. Probabilistic classes cover BPP (BPP $\subseteq$ P/poly\cite{valiant1979complexity}), PP, and \#P (matrix permanent is \#P-complete\cite{toda1991pp}). Oracle-defined PH is in PSPACE; theorems link these classes and reveal hierarchy influences\cite{sipser1983complexity,karp1980some}.

\subsection{Quantum Computation Model and Complexity}

Quantum computation models and complexity have revolutionized computational science, fundamentally reshaping academic understanding of computability and computational complexity. Quantum computation models have greatly expanded our understanding of computability, as quantum computing holds enormous application potential in specific areas where classical computing faces significant difficulties.

\subsubsection{Types of Graphics Quantum Computation Model}

The 1980s marked the initial phase of quantum computation, characterized by the development of physical and mathematical models. Paul Benioff proposed the quantum Turing machine in 1980\cite{benioff1980quantum}. Richard Feynman proposed quantum simulators for efficient quantum system simulation in 1982\cite{feynman1982simulating}. David Deutsch introduced the universal quantum computer in 1985\cite{deutsch1985quantum} and the quantum circuit model in 1989\cite{deutsch1989networks}. From the 1990s to 2000s, quantum models were refined and standardized. Chi-Chih Yao proved the equivalence of quantum Turing machines and quantum circuits, defining their complexity metrics in 1993\cite{yao1993quantum}. Seth Lloyd and Samuel L. Braunstein proposed the continuous-variable model in 1998\cite{lloyd1999continuous}. Edward Farhi et al. put forward the quantum annealing model in 2000\cite{farhi2000quantum}. Robert Raussendorf and Hans J. Briegel proposed the measurement-based model in 2000\cite{raussendorf2001oneway}. Alexei Kitaev proposed the topological model in 2003\cite{kitaev2003fault}. Since the 2000s, advances in hardware (superconducting, ion trap, photonic) have spurred new models, such as the measurement-based model in 2001, relying on entangled cluster states and selective measurements, is well-suited for optical quantum computation\cite{raussendorf2003measurement}.

The Quantum Turing Machine (QTM) is one of the key models in quantum computation complexity theory. QTM integrates the deterministic computation framework of TM with quantum superposition states and unitary evolution, and QTMs are primarily categorized into the following types, which form a hierarchical relationship\cite{nielsen2010quantum}.

Generalized Quantum Turing Machine (GQTM). As the most comprehensive model, it accounts for non-terminating branches of quantum states, allowing computation processes to evolve through superposition states and thus possessing the strongest computational power.

Modular Quantum Turing Machine (MQTM). A subset of GQTM, typically referring to models under specific constraints, whose computational power does not exceed that of GQTM.

Randomized Quantum Turing Machine (ROQTM). A further constrained model within MQTM, with the weakest computational power. These models were proposed to mathematically formalize the classification and capabilities of quantum computers in solving specific problems.

QTM retains the core structure of TM (such as the state machine, read-write head, and tape) while elevating the machine’s state space and transition rules to Hilbert space, and introducing the constraints of ``amplitude normalization and unitarity''. A QTM is typically defined as a seven-tuple (or more tuples, depending on the specific definition). 

\begin{equation}M_Q = (Q, \Sigma, \Gamma, \delta, q_0, B, F)\end{equation}
\subsubsection{Quantum Complexity}
In quantum computation theory, complexity is defined as spatiotemporal complexity, specifically the number of tape units (space) required to execute all branch computations on a QTM and the number of steps (time) needed to reach the final configuration\cite{lipton2010savitch}. The quantum computation complexity is primarily measured by the following metrics.
\begin{enumerate}
	\item Circuit Complexity. The minimum number of gates(maximum sequence length of parallel quantum gates) required to achieve a specific state transformation, typically involving the completeness of a universal gate set and the approximation problem addressed by the Solovay–Kitaev theorem.
	\item Space Complexity. The total number of quantum bits required to achieve a specific state transformation.
	
	\item Query Complexity. In the oracle model, the minimal upper bound on the number of queries a computer makes to an oracle. This is a core metric highlighting the difference between quantum and classical computational complexity (e.g., the quadratic acceleration achieved by Grover’s algorithm)\cite{yan2015classical}.
	\item Entanglement Measure. In certain models, generating sufficient entangled states (e.g., GHZ states) is a prerequisite for achieving specific state transformations, making the complexity of entanglement generation an emerging key metric\cite{ma2025multipartite}.
\end{enumerate}

Quantum computational complexity focuses on the complexity class BQP, which describes the computational power of bounded-error quantum polynomial-time algorithms. Although quantum computation leverages unitary evolution and quantum properties for parallel information processing, it does not extend the set of computable functions compared with classical computation, only reducing complexity exponentially or quadratically for specific problems. The inclusion relation P$\subseteq$BPP$\subseteq$BQP$\subseteq$PP$\subseteq$PSPACE holds, meaning quantum polynomial-time computations can be classically simulated with polynomial space\cite{bernstein1993quantum}. As a quantum generalization of NP, QMA satisfies NP$\subseteq$MA$\subseteq$QMA$\subseteq$PP\cite{kempe2004complexity}. Core theorems such as BQP$\subseteq$PP and QMA$\subseteq$PP reveal the limits of quantum computing power.

\section{The Hybrid Computation Model HTM}
The integration of hybrid computation has emerged as a new computation paradigm. However, the absence of theoretical models has hindered progress in task flow models and resource scheduling mechanisms. Thus, HTM elevates the field from empirical engineering practices to standardized theoretical guidance.

\subsection{Formal Description of the HTM}
\begin{figure}[!htb]
	\centering
	\includegraphics[width=0.98\linewidth]{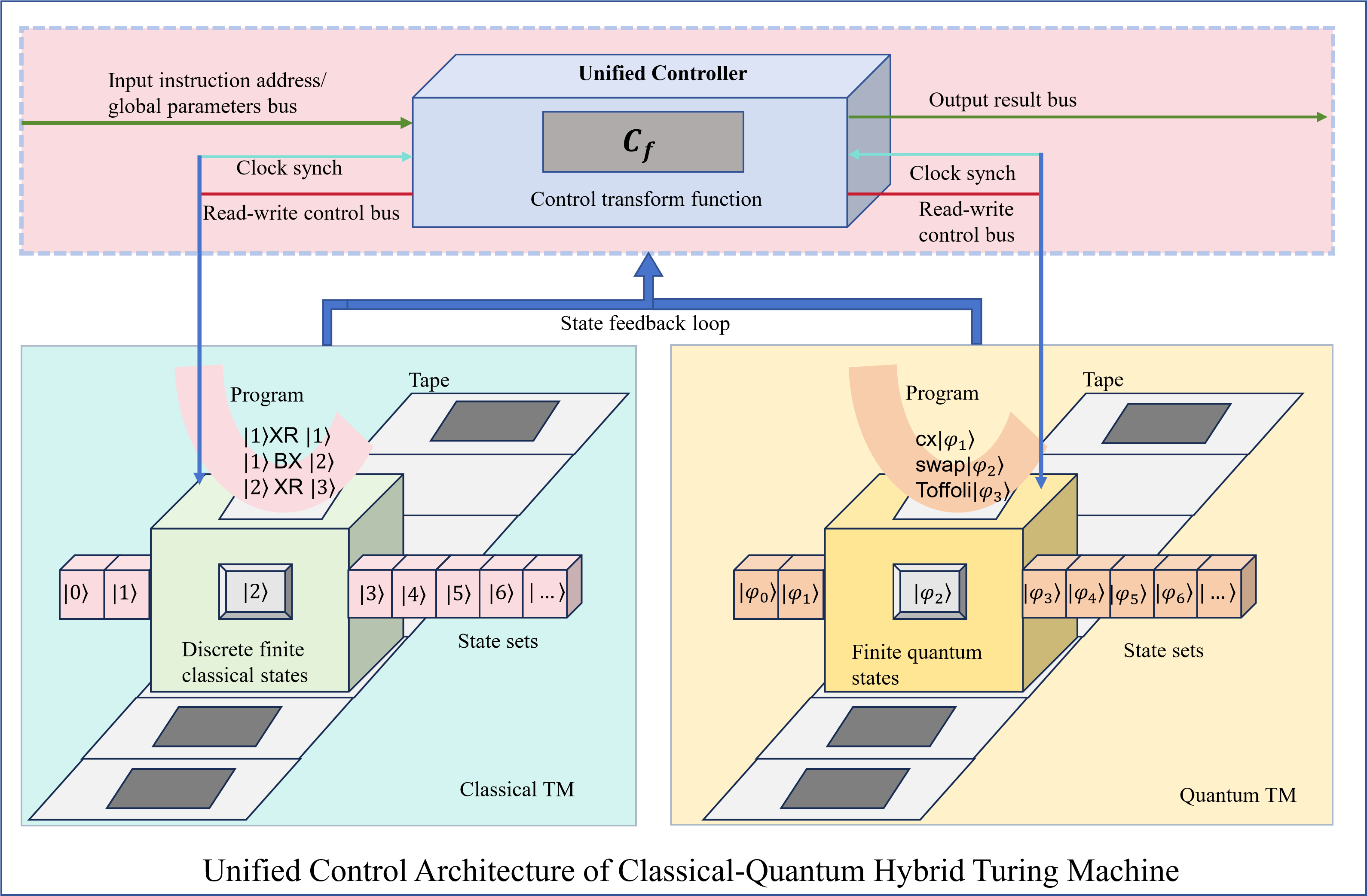}
	\caption{Hybrid Turing Machine Architecture: Unified Control with Dual Read-Write Heads.}
	\label{fig:1}
\end{figure}

HTM resolves the disconnect between classical and quantum computation in control logic and resource scheduling to achieve deep synergy and optimal resource allocation. First, the research formalizes unified models based on ``Turing Machine theory'', bridging the gap between empirical engineering and theoretical standardization in computation field (as shown in \cref{fig:1}).

Based on the schematic diagram of the Hybrid Turing Machine (HTM) shown in \cref{fig:1}, together with the 7‑tuple formal definition, the mathematical description of the state set and transition function, and the three‑layer functional extension provided in the text, this section systematically expands and elaborates the architecture, state space, transition mechanism, and control‑sharing principle of the HTM. The goal is to reveal the deep integration mechanism between classical and quantum computation under a unified controller.

\subsubsection{Overall Structure of the HTM: Dual Read/Write Heads and a Unified Controller} 

As illustrated in \cref{fig:1}, the HTM introduces two independent read/write heads, a classical head and a quantum head on the basis of the classical Turing machine and the quantum Turing machine. The two heads share the same tape (which is physically or logically divided into a classical track and a quantum track) but are both scheduled and controlled by a single unified controller($C_f$). The classical head processes binary symbols (0/1) following deterministic state transitions; the quantum head manipulates quantum states by performing unitary transformations and measurements. The current state of the controller, the transition rules, and the program instructions jointly determine whether a classical operation, a quantum operation, or a coordinated operation is executed in each step.

This dual‑head structure breaks the limitation of a single control flow in traditional Turing machines, allowing the HTM to describe both classical algorithms and quantum algorithms within the same computational framework and to support their interleaved execution.

\subsubsection{7‑Tuple Formal Definition of the HTM}

Referring to Table~\ref{Table:1} in the original text, the HTM is defined as a seven-tuple \( \text{HTM} = (Q, \Sigma, \Gamma, \delta, q_0, B, F) \) to define HTM.
\begin{table}[htbp]
	\centering
	\caption{HTM Symbol and Name Interpretation}   
	\label{Table:1}                      
	\begin{tabular}{c c p{5cm}}
		\toprule
		\textbf{Symbol} & \textbf{Name} & \textbf{Interpretation} \\
		\midrule
		$Q$       & Finite state set    & Finite non-empty set \\
		$\Sigma$  & Input alphabet      & Input symbol set \\
		$\Gamma$  & Tape alphabet       & The set of tape symbols \\
		$\delta$  & Transition function & $\delta: Q \times \Gamma \rightarrow Q \times \Gamma \times \{L, R, H\} \times C$ \\
		$q_0$     & Initial state       & Initial state of the controller \\
		$B$       & Blank symbol        & The ground state of the tape lattice $U$ \\
		$F$       & Termination state set & Calculation termination, input state rejected or accepted \\
		\bottomrule
	\end{tabular}
\end{table}
\subsubsection{Mathematical Description of the State Set Q} 

Let \( Q = \{\ket{\phi_0}, \ket{\phi_1}, \dots\} \) encompasses both classical states and quantum states. Taking a single qubit as an example. For a single-bit quantum state
\begin{equation}
	\ket{\phi_q} = \alpha \ket{0} + \beta \ket{1}
	\label{eq:quantum_state}
\end{equation}
where the coefficients satisfy the normalization condition \(|\alpha|^2 + |\beta|^2 = 1\); When the system operates in a purely classical mode, only one coefficient is nonzero (equal to 1), which reduces to a deterministic state. This description unifies the state representation of a classical Turing machine (where probability amplitudes degenerate to a special case of discrete probabilities) and that of a quantum Turing machine (which allows arbitrary complex amplitudes). For a single-bit classical state
\begin{equation}
	\ket{\phi_c} = \ket{0} \quad \text{or} \quad \ket{\phi_c} = \ket{1}
	\label{eq:classical_state}
\end{equation}
which demonstrates that classical states form a special case of quantum states.
For the classical tape, symbols from the input alphabet \( \Sigma \) and tape alphabet \( \Gamma \) (denoted by \( \alpha, \beta \)) are restricted to binary values 0 and 1, with the constraint that \( \alpha \) and \( \beta \) cannot be identical.

For the quantum tape, \( \alpha, \beta \) range continuously over \([0,1]\) and satisfy \(|\alpha|^2 + |\beta|^2 = 1\).
\subsubsection{Unitary Realization of the Transition Function $\delta$}

Each step of the transition function \(\delta\) is implemented by a unitary operator \(U\). For a quantum state,
\begin{equation}
	\delta = \{U \mid U^\dagger U = I\}
	\label{eq:unitary}
\end{equation}
The result of \(U\) acting on the state is as follows:
\begin{equation}
	U \ket{\varphi_q} = \alpha' \ket{0} + \beta' \ket{1}
	\label{eq:U_action}
\end{equation}
For a classical state, which is a subset of quantum states, the same operation applies, but the result must be guaranteed to still belong to the classical state set (i.e., a permutation of computational basis states).
\begin{equation}
	U \ket{\varphi_c} : \ket{0} \mapsto \ket{1} \text{ or } \ket{1} \mapsto \ket{0}
	\label{eq:classical_action}
\end{equation}

The HTM adopts deterministic state transitions. Given the current state and the symbol read, the next state is uniquely determined. Let the initial state be \(q_0\). Computation stops when the machine enters a terminal state \(q_f \in F\).

A homomorphic mapping between classical Boolean logic operations and quantum gate operations is established as Table~\ref{Table:2}.
\begin{table}[htbp]
	\centering
	\caption{Homomorphic Mapping Between Classical Boolean Logic Operations and Quantum Operations}
	\label{Table:2}
	\begin{tabular}{c c c}
		\toprule
		\textbf{Classical Logic} & \textbf{Quantum Gate} & \textbf{Operation Definition} \\
		\midrule
		AND   & Toffoli gate & Toffoli \\
		OR    & CNOT gate    & CNOT \\
		NOT   & X gate       & X \\
		\bottomrule
	\end{tabular}
\end{table}

Through this mapping, any classical Boolean circuit can be transformed into a unitary network composed of quantum gates, thereby enabling the HTM to execute classical control flow and quantum data flow in a unified manner.
\subsubsection{Three‑Layer Functional Logic Extension}

To achieve control sharing and resource scheduling, the original text proposes a three‑layer functional extension of the HTM. The following subsections elaborate on each layer in conjunction with the structure shown in \cref{fig:1}.

\paragraph{Decision and Control Layer}

This layer is implemented by the core decision module
of the unified controller \( C_f \). It decomposes the overall
computational task \( T \) into two subtasks:$C_f : T \mapsto \{T_c, T_q\}\label{eq:task_decomp}$.

where \( T_c \) represents classical control tasks, including
protocol handshaking, instruction scheduling, branch jumps,
loop control, etc.; \( T_q \) represents quantum communication
and quantum computation tasks, including entanglement
preparation, quantum state transmission, superdense coding,
execution of quantum gate sequences, etc.

The decomposition is based on task parallelism and resource dependencies. The controller dynamically adjusts the ratio and execution order of $T_c$ and $T_q$ through a predefined decision‑control function $f_{ctrl}$, thereby achieving adaptive resource scheduling. For example, in a variational quantum algorithm, $T_c$ performs parameter updates and convergence checking, while $T_q$ executes repeated sampling measurements of the quantum circuit.

\paragraph{State Transition and Protocol Mapping Layer}
This layer is responsible for converting classical control instructions into low‑level quantum physical operations. The core requirement is that all state transition operations $\delta$ must preserve unitarity to ensure probability conservation in physical processes.

Specifically, this layer maintains a protocol mapping table (e.g., Table~\ref{Table:2} in the original text). For a logical expression in the classical control flow, such as ``if(a AND b) then ...'', the middle layer compiles it into a Toffoli-gate control sequence; for an ``OR'' logic, it is compiled into a combination of CNOT gates. This layer also translates read-write head movement commands (L, R, H) from the controller into tape address offsets or quantum register index remappings.

\paragraph{Hybrid State Space Layer}
The bottom layer defines the physical state space Q of the HTM. Unlike the discrete bits stored in a classical Turing machine, the state space of the HTM is the tensor product space of a set of basis states. At any time, the system may reside in a linear superposition of these basis states, subject to the core constraint of probability amplitude normalization.

The classical read-write head directly reads and writes the computational basis vectors (i.e., classical states) in this space, while the quantum head can perform unitary transformations or projective measurements on arbitrary superposition states. The tape itself is partitioned into a classical track (storing binary symbols) and a quantum track (storing quantum state amplitudes), with coordinated access managed by the unified controller’s address management mechanism.

In summary, based on \cref{fig:1} and the above formal description, the HTM achieves the following key breakthroughs:

Unified formal foundation. The HTM incorporates both the classical Turing machine and the quantum Turing machine into a single 7‑tuple framework. In this framework, the state set and the transition function are described using unitary operators, which mathematically reveals that classical computation is merely a restriction of quantum computation to the computational basis. This unification bridges the long‑standing gap between classical and quantum theoretical models.

Shared control logic. A single unified controller Cf manages both the classical and quantum read-write heads, thereby avoiding the redundancy and potential conflicts that would inevitably arise if each heterogeneous computing unit maintained its own independent control state machine. Moreover, the program stored in the controller contains both classical instructions (e.g., ``XR'', ``BX'') and quantum gate instructions (e.g., controlled‑X, swap, Toffoli), achieving a truly unified instruction set for hybrid computation.

Adaptive resource scheduling. Through the three‑layer functional logic architecture, the HTM can dynamically allocate the proportion of classical control tasks to quantum computation tasks based on the specific type of the overall task. This adaptive capability is especially suitable for hybrid algorithms such as variational quantum eigensolvers (VQE) and quantum approximate optimization algorithms (QAOA), which require frequent alternation between classical parameter optimization and quantum circuit sampling.

Scalability and verifiability. Because all state transition operations are required to preserve unitarity, the entire evolution of the HTM is reversible and probability‑conserving. This property not only respects the fundamental laws of quantum mechanics but also provides a solid foundation for the formal verification of correctness and for rigorous complexity analysis of the HTM, making it highly scalable to larger systems and more complex hybrid tasks.
\subsection{Computation Complexity Analysis}
This study conducts computational complexity analysis within the paradigms of both classical and quantum computation models. By decomposing the logical steps of the algorithm into elementary operations, it achieves compatibility and unification between HTM and the existing complexity analysis paradigm.
\subsubsection{Complexity Analysis Methods}
State transformation lays an operational foundation for the quantification, decomposition, and verification of HTM complexity. Thus, complexity analysis guides the scheduling and allocation of classical and quantum micro-operations, as well as enables accurate quantification and decomposition verification of hybrid models, while directly guides the scheduling and resource allocation, providing strong theoretical support for the engineering implementation and performance optimization.

The complexity description employs the Dirac notation system, where \( |\varphi_i\rangle \) and \( |\varphi_j\rangle \) denote the system's state at discrete time step \( i \) and \( j \) (\( j > i \)). The state evolution follows the unitary transformation principle, driven by the unitary operator \( U \). The evolution equation can be expressed as\[|\varphi_j\rangle = U |\varphi_i\rangle,\]
where\[|\varphi_i\rangle = \sum_{i=0}^{N-1} \alpha_i |\alpha_i\rangle,\]\[|\varphi_j\rangle = \sum_{j=0}^{N-1} \beta_j |\beta_j\rangle.\]\( |\alpha_i\rangle \)and \( |\beta_j\rangle \) are computation bases of the Hilbert space, and \( \alpha_i \) and \( \beta_j \) denoting bases' amplitudes respectively. Unitarity is the constraint governing quantum evolution, ensuring the normalization of quantum states and the reversibility of evolution.
Based on quantum entanglement(ent.) and superposition(sup.), the state evaluation can be categorized into nine distinct scenarios (as shown in Table~\ref{Table:3}).
\begin{table}[htbp]
	\centering
	\setlength{\tabcolsep}{6pt} 
	\caption{Classification of Quantum State Evolution Properties (The evaluation from \( |\phi_i\rangle \) to \( |\phi_j\rangle \))}
	\label{Table:3}
	\begin{tabularx}{\columnwidth}{p{6cm} p{6cm}} 
		\toprule
		\textbf{States entanglement properties} & \textbf{States superposition properties} \\
		\midrule
		\multirow{1}{=}{ent. $\to$ ent.} & $\bullet$ sup. $\to$ sup. \\
		\hline
		\multirow{1}{=}{Non-ent. $\to$ ent.} & $\bullet$ Non-sup. $\to$ sup.\\& $\bullet$ sup. $\to$ sup. \\
		\hline
		\multirow{2}{=}{ent. $\to$ Non-ent.} & $\bullet$ sup. $\to$ Non-sup. \\
		& $\bullet$ sup. $\to$ sup. \\
		\hline
		\multirow{3}{=}{Non-ent. $\to$ Non-ent.} 
		& $\bullet$ Non-sup. $\to$ Non-sup. \\
		& $\bullet$ sup. $\to$ Non-sup. \\
		& $\bullet$ Non-sup. $\to$ sup. \\
		& $\bullet$ sup. $\to$ sup. \\
		\bottomrule
	\end{tabularx}
\end{table}
\subsubsection{Computation Complexity Results}
Based on the entanglement and superposition properties of the initial and target states, this study analyzes the time and space complexity of realizing this state transformation process in both classical and quantum computation, where the upper bound of the complexity is attained. The conclusions are presented in Table~\ref{Table:4}.
\begin{sidewaystable}
	\centering
	\small
	\setlength{\tabcolsep}{4pt}
	\caption{Classification of State Transformation Properties}
	\label{Table:4}
	\begin{tabularx}{\textheight}{   
			p{2.8cm}   
			p{0.7cm}   
			X          
			p{2.8cm}   
		}
		\toprule
		\multicolumn{1}{c}{\multirow{2}{*}{n-bit state transformation \( |\varphi_i\rangle \mapsto |\varphi_i\rangle \)}} 
		& \multicolumn{3}{c}{Complexity} \\
		\cmidrule{2-4}
		&   & \multicolumn{1}{c}{Classic (tape length \( n2^n \))} 
		& \multicolumn{1}{c}{Quantum (tape length \(n\))} \\
		\midrule
		\multirow{2}{*}{Non-ent.\ $\rightarrow$\ Non-ent.}
		& Time
		& \multirow{2}{*}{
			\parbox{\linewidth}{%
				\begin{enumerate}
					\setlength{\itemsep}{0pt}
					\setlength{\parskip}{0pt}
					\item \( O(cn) \), \( c \leq 2^n \), \( c \) = number of sup. basis states  
					\item \( O(n) \), no sup. state involved
				\end{enumerate}
			}%
		}
		& \(O(cn)\), \( c \leq 2^n \) \\
		\cmidrule{2-2}\cmidrule{4-4}
		& Space & & \(n\) \\
		\midrule
		\multirow{2}{*}{
			\parbox{\linewidth}{%
				\(\text{ent.} \rightarrow \text{Non-ent.} /\)\\
				\(\text{Non-ent.} \rightarrow \text{ent.}\)
			}%
		}
		& Time
		& \multirow{2}{*}{
			\parbox{\linewidth}{%
				\(O(cn), c \leq 2^n\), \(|\varphi_i\rangle\) or \(|\varphi_j\rangle\) is sup. state.
			}%
		}
		& \(O(cn)\), \(c \leq n\) \\
		\cmidrule{2-2}\cmidrule{4-4}
		& Space & & \(n\) \\
		\midrule
		\multirow{2}{*}{\(\text{ent.} \rightarrow \text{ent.}\)}
		& Time
		& \multirow{2}{*}{
			\parbox{\linewidth}{%
				\(O(cn), c \leq 2^n\), \(|\varphi_i\rangle\) and \(|\varphi_j\rangle\) are sup. states,\\
				\(c\) = max number of sup. basis states in both.
			}%
		}
		& \(O(cn)\), \( c \leq 2^n \) \\
		\cmidrule{2-2}\cmidrule{4-4}
		& Space & & \(n\) \\
		\bottomrule
		\addlinespace
		\multicolumn{4}{l}{\footnotesize\textit{Note:} ent. = entangled, Non-ent. = non-entangled, sup. = superposition.}
	\end{tabularx}
\end{sidewaystable}
Due to the differences between classical and quantum computing paradigms, for time complexity metrics, classical computation adopts the number of Turing machine steps (including read, write and move operations), while quantum computation uses the number of single- and two-qubit elementary gate operations.
For the non-ent. to non-ent. case, quantum time complexity is exponentially superior to the classical one, whereas the space complexity of both is polynomial. For other cases, the worst-case time complexity of both classical and quantum computation is exponential, but quantum space complexity is exponentially better than the classical one.

The above conclusions are illustrated with an example in the following. We focus solely on the ``Non-ent.→ Non-ent.'' scenario for complexity analysis.
\paragraph{Non-sup. state to non-sup. state}
In this case, \( |\varphi_i\rangle = |\alpha_i\rangle \), \( |\varphi_j\rangle = |\beta_j\rangle \), \( |\varphi_i\rangle\)and\( |\varphi_j\rangle\) represent non-sup. states, consistent with classical data representation.

Unitary operator \( U = U_1 \otimes U_2 \otimes \cdots \otimes U_n \), where each \( U_i \) is a single-qubit gate. Moreover, if \( U_i \) has an equivalent single-bit classical logic gate (such as the X gate), then the corresponding complexity \( U_C \) is the same for both classical and quantum cases.

\begin{equation}
	|\varphi_i\rangle = U |\varphi_i\rangle, \quad \beta_i = U_c * \alpha_i 
\end{equation}
For example, \( |0101\rangle = X^{\otimes 4} |1010\rangle, 0101=\neg 1 \land \neg 0 \land \neg 1 \land \neg 0 \).Both classical and quantum NOT operations necessitate four computational steps and four bits of spatial resources.
\paragraph{Superposition state to non-sup. state}
\begin{equation}
	|\varphi_i\rangle = \sum_{i=0}^{N-1} \alpha_i |\alpha_i\rangle, \quad |\varphi_j\rangle = |\beta_j\rangle 
\end{equation}

For quantum,
\[
|\varphi_j\rangle = U |\varphi_i\rangle, \quad U = U_1 \otimes U_2 \otimes \cdots \otimes U_n,
\]
this process requires at most \( n \) gates, with circuit depth is 1 and space complexity is \( O(n) \).

For classical systems, the time complexity equals the number of steps for the read-write head to traverse the tape, which is \( 2^n \), and the space complexity is \( O(2^n) \).
\paragraph{Non-sup. state to sup. state}
The complexity analysis for this process is consistent with that of the ``sup. state to non-sup. state'' scenario.
\paragraph{Sup. state to sup. state}
\begin{equation}
	|\phi_i\rangle = \sum_{i=0}^{N-1} \alpha_i |\alpha_i\rangle, \quad |\phi_j\rangle = \sum_{j=0}^{N-1} \beta_j |\beta_j\rangle 
\end{equation}

For quantum, \(|\phi_i\rangle = U|\phi_i\rangle\). Sup. without ent. can be realized via single-qubit gates, requiring at most \(n\) gates, circuit depth 1, and space complexity \(O(2^n)\).

For classical systems, non-ent. and sup. states correspond to unitary transformations on \(2^n\)-dimensional vector. The time complexity is \(O(2^n)\) (tape traversal steps of the read-write head), and the space complexity is \(O(2^n)\).

In summary, owing to quantum characteristics, quantum computing exhibits an exponential advantage in space complexity. Regarding time complexity, an exponential quantum advantage exists only in the case of Non-ent. $\to$ Non-ent., and the worst-case time complexity is equivalent in all other scenarios.

\subsection{The Control-sharing Mechanism}
Control sharing is a key feature of the Hybrid Computing Model (HTM). It employs a unified controller ($C_f$) to achieve coordinated scheduling and deep integration of classical and quantum computing resources. This paper systematically describes the data and control interaction mechanisms among the input/output units, the memory unit, the unified controller, and the two computing models. The analysis emphasizes the strong bidirectional collaboration in the classical computing model and the asymmetric, unidirectional control‑dominant design for the quantum computing model. While reducing redundant control logic in heterogeneous units, the control‑sharing mechanism preserves the execution efficiency of each computing paradigm, thereby providing theoretical support for hybrid computing architecture design.

When an external input (computing tasks, initial data, control commands) arrives,$C_f$ first parses the task, splits it into parallelizable subtasks, and decides which parts should be executed by the classical TM and which by the quantum TM. $C_f$ manages all data uniformly through the memory unit: input data, classical intermediate results, pre‑measurement quantum states, and final outputs. The memory uses either unified addressing or partitioned management to simultaneously satisfy the classical TM’s need for frequent random access and the quantum TM’s need for sequential/batch read/write.
\begin{enumerate}
\item For the classical TM, the interaction is bidirectional and strongly collaborative. $C_f$ sends downlink commands to the classical TM (register settings, task start/stop, runtime parameter adjustments). During execution, the classical TM can feed intermediate results or measurement data (originating from the quantum TM) back to $C_f$ or write them directly to memory. This uplink data path enables dynamic feedback: for example, in the Variational Quantum Eigensolver (VQE) algorithm, the classical optimizer must repeatedly read quantum measurement outcomes and update parameters, then send the new parameters back to the quantum TM via $C_f$. Hence, under control sharing, the classical TM retains its full bidirectional control and data capability.

\item For the quantum TM, the interaction is fundamentally different. It is asymmetric and control‑dominant, with a unidirectional control flow. Due to the physical constraints of current quantum hardware (short coherence times, need for pre‑scheduled operation sequences, high sensitivity to feedback latency), $C_f$ adopts a unidirectional, control‑dominant design. All control signals – task scheduling, configuration parameters, execution commands, error‑correction timing, pulse generation – flow one way from $C_f$ to the quantum TM. The quantum TM never actively initiates a control request; it passively receives and executes commands. Measurement results can be sent back to $C_f$  or memory through a separate data channel, but the control stream remains strictly from $C_f$ to the quantum TM. This creates a pattern of ``control commands strictly downstream, measurement data limited upstream''. Complex real‑time decisions (e.g., dynamic decoupling, conditional reset) are handled by $C_f$ as part of the shared control logic, while the quantum TM focuses only on low‑latency measurement feedback and does not need to maintain its own full control state machine.

There will be three major benefits arise from this control‑sharing design:
\begin{itemize}
	\item Elimination of redundant control logic. Neither the classical TM nor the quantum TM needs to implement common control functions (task queuing, resource arbitration, memory mapping, etc.). All shared control logic is centralized in Cf, drastically reducing hardware/firmware duplication across heterogeneous units.
	\item Preserved execution efficiency. The classical TM still enjoys high-bandwidth bidirectional interaction, making it efficient for data‑intensive workloads. The quantum TM receives a streamlined unidirectional control interface, avoiding extra control overhead that would otherwise compromise the precise timing of quantum operations.
	\item Physical constraint compatibility and scalability. When adding more classical or quantum computing units, developers only need to comply with the standard signal interfaces defined by $C_f$; no modification of internal control state machines inside existing units is required.
\end{itemize}

\end{enumerate}

The unified controller allocates control responsibilities differently according to the distinct natures of the classical TM and the quantum TM. The classical side retains strong bidirectional collaboration to support dynamic feedback loops in hybrid algorithms. The quantum side adopts a unidirectional, control‑dominant interface to respect the inherent physical constraints of quantum hardware. By sharing the same controller, the system reduces heterogeneous redundancy while achieving deep integration of two very different computing resources – this is the theoretical foundation and structural reference for efficient hybrid computing architecture design.

\section{An Example Illustration of HTM}

\subsection{Problem and HTM Instantiation (No Resource Constraints)}

\textbf{Problem Background:} Consider a 5-vertex cycle graph
\( G = (V, E) \), where \( V = \{0, 1, 2, 3, 4\} \) and \( E =\{(0, 1), (1, 2), (2, 3), (3, 4), (4, 0)\} \). QAOA with depth \( p=1 \) is used to solve Max-Cut. The quantum circuit includes initial state preparation, the problem unitary operator \( U_c(\gamma) = e^{-i\gamma H_c} \), and the mixing unitary operator \( U_M(\beta) = e^{-i\beta H_M} \),

where
\[
H_c = \frac{1}{2} \sum_{(i,j) \in E} (I - Z_i Z_j), \quad H_M = \sum_{i=0}^4 X_i.
\]

\textbf{Objective:} To formally describe the complete execution of this hybrid algorithm within the HTM framework, including the alternation between classical parameter updates and quantum circuit evaluation, scheduling by the unified controller, operational transformations, and complexity analysis.

\textbf{HTM Construction:} Construct HTM $M_{QAOA}$=(Q, $\Sigma$, $\Gamma$, $\delta$, $q_0$, B, F), where:

\begin{itemize}
	\item State set \( Q = Q_{\text{class}} \cup Q_{\text{quantum}} \)
where
	\[
	Q_{\text{class}} = \{ q_{\text{int}}, q_{\text{param}}, q_{\text{prep}}, q_{\text{applyC}}, q_{\text{applyM}}, q_{\text{sample}}, q_{\text{update}}, q_{\text{term}} \}
	\]
	\[
	Q_{\text{quantum}} = \left\{ |\psi\rangle = \sum_{x \in \{0,1\}^5} \alpha_x |x\rangle \right\}
	\]
	
	\item Tape alphabet \( \Gamma = \Gamma_{\text{class}} \cup \Gamma_{\text{quantum}} \)
where
	\[
	\Gamma_{\text{class}} = \{ 0,1, \sqcup \} \cup \{ \gamma, \beta, \text{Ks}, \text{$E_{best}$}, \text{m} \}
	\]
	\[
	\Gamma_{\text{quantum}} = \{ |x\rangle \mid x \in \{0,1\}^5 \} \cup \{ \text{all superposition states} \}
	\]
	
	\item Initial state \( q_0 = q_{\text{int}} \)
	
	\item blank symbol \( B = (\sqcup, |0\rangle^{\otimes 5}) \)
	
	\item Final state set \( F = \{ q_{\text{term}} \} \)
\end{itemize}

The tape is divided into a classical track (storing parameters, counters, measurement results) and a quantum track (storing the 5‑bit quantum state). The classical read-write head and the quantum read/write head share the same physical address space on the tape and are coordinated by a unified controller $C_f$ through an address management mechanism.

\subsection{Hybrid Task Decomposition and Formal Transition Sequence}
The controller decomposes the overall computational task T into a classical control subtask $T_c$ and a quantum computation subtask $T_q$, and defines a decision function $f_{ctrl}$ for dynamic scheduling. Assume the maximum number of outer QAOA iterations is $K_{max}$=100. The complete transition sequence for a single iteration (given parameters $\gamma$,$\beta$) is provided below.

Assume that initially the classical track stores \( \gamma = 0.6 \), \( \beta = 0.4 \),  
\( k = 0 \), \( E_{\text{best}} = \infty \), and the initial state of the quantum track is \( |0\rangle^{\otimes 5} \).

\begin{table}[htbp]
	\centering
	\footnotesize                     
	\setlength{\tabcolsep}{4pt}       
	\caption{Classification of State Transformation Properties}
	\label{Table:5}
	\begin{tabularx}{\columnwidth}{
			>{\centering\arraybackslash}p{0.6cm}   
			>{\centering\arraybackslash}p{1.0cm}   
			>{\centering\arraybackslash}p{1.5cm}   
			X
			>{\centering\arraybackslash}p{2cm} 
		}
		\hline
		\textbf{Step} & \textbf{Controller State} & \textbf{Activated Component} & \textbf{Operation} & \textbf{Formal Transition$\delta$} \\
		\hline
		0    & \( q_{\text{int}} \)    & Classical read-write head & Read parameters and iteration count & \( \delta(q_{\text{int}}, \sqcup) \) \\
		\hline
		1    & \( q_{\text{param}} \)  & Classical read-write head & Check \( k < k_{\max} \); if true, continue & \( \delta(q_{\text{param}}, k) \) \\
		\hline
		2    & \( q_{\text{prep}} \)   & Quantum read-write head & Prepare \( |+\rangle^{\otimes 5} \) & \( \delta(q_{\text{param}}, |+\rangle^{\otimes 5}) \) \\
		\hline
		3    & \( q_{\text{apply}} \)  & Quantum read-write head & Apply \( U_c(\gamma) \) & \( \delta(q_{\text{apply}}, |+\rangle^{\otimes 5}) \) \\
		\hline
		4    & \( q_{\text{applyM}} \) & Quantum read-write head & Apply \( U_M(\beta) \) & \( \delta(q_{\text{applyM}}, |\psi_c\rangle) \) \\
		\hline
		5    & \( q_{\text{sample}} \) & Quantum read-write head & Measure all qubits in the Z basis, obtaining \( m \in \{0,1\}^5 \) & \( \delta(q_{\text{sample}}, |\psi_{\text{final}}\rangle) \) \\
		\hline
		6    & \( q_{\text{update}} \) & Classical Head & Calculate energy \( E(m) \); if \( E(m) < E_{\text{best}} \), update \( E_{\text{best}} \); update \( (\gamma, \beta) \) based on gradient information; $k \leftarrow k+1$ & \( \delta(q_{\text{update}}, m, E_{\text{best}}) \) \\
		\hline
	\end{tabularx}
\end{table}
Where
\begin{flalign*}
	& E(m) = \frac{1}{2} \sum_{(i,j)\in E} \bigl(1 - (-1)^{m_i\oplus m_j}\bigr), &\\
	& \delta(q_{\text{int}}, \sqcup) = \bigl(q_{\text{param}}, (\gamma, \beta, k), H, \text{null}\bigr), &\\
	& \delta(q_{\text{param}}, k) = \bigl(q_{\text{param}}, \text{true}, R, \text{comp}(k, K_{\max})\bigr), &\\
	& \delta(q_{\text{param}}, |0\rangle^{\otimes 5}) = \bigl(q_{\text{applyC}}, |+\rangle^{\otimes 5}, H, H^{\otimes 5}\bigr), &\\
	& \delta(q_{\text{applyC}}, |+\rangle^{\otimes 5}) = \bigl(q_{\text{applyM}}, |\psi_C\rangle, H, U_C(\gamma)\bigr), &\\
	& \delta(q_{\text{applyM}}, |\psi_C\rangle) = \bigl(q_{\text{sample}}, |\psi_{\text{final}}\rangle, H, U_M(\beta)\bigr), &\\
	& \delta(q_{\text{sample}}, |\psi_{\text{final}}\rangle) = \bigl(q_{\text{update}}, m, R, \text{Measure}(Z^{\otimes 5})\bigr), &\\
	& \delta(q_{\text{update}}, m, E_{\text{best}}) = \bigl( \text{if } k < k_{\max} \text{ then } q_{\text{param}} ,\text{else } q_{\text{term}}, (\gamma', \beta', k'), L, \text{classical\_circuit}\bigr). &
\end{flalign*}

The \( H^{\otimes 5} \) in step 2 is the tensor product of Hadamard gates, an operation of the quantum head.

The \( U_c(\gamma) \) in step 3 is mapped to a sequence of quantum gates via an intermediate layer protocol. Since \( H_c \) is a sum of ZZ interactions, \( \exp(-i\gamma H_c) \) can be decomposed into
\[
\prod_{(i,j) \in E} \exp\bigl(-i\gamma (I - Z_i Z_j)/2\bigr).
\]
Each factor \( \exp(i\gamma (Z_i Z_j)/2) \) is compiled as: \( \text{CNOT}(i,j) \, RZ(\gamma, j) \, \\ \text{CNOT}(i,j) \). The intermediate layer maintains a protocol mapping table, linking the physical operator to its gate-level implementation.

The \( U_M(\beta) = \bigotimes_{i=0}^{4} R_x(2\beta)_i \) in step 4 is directly mapped to 5 parallel \( R_x \) gates.

The classical\_circuit in step 6 is implemented via homomorphic mapping (Toffoli/CNOT networks) to realize: comparators, adders, and multipliers (for gradient descent). This circuit is executed entirely on the classical track by the classical head and does not involve the quantum head.
\subsection{Operational Transformation and Homomorphic Mapping Example}
\subsubsection{Quantum Operation Compilation}
For edge \((i, j)\), the intermediate layer converts \(\exp(i\gamma(Z_iZ_j)/2)\) into the following executable instruction sequence for HTM (written into the quantum head's microcode):
\begin{center}
	CNOT \( i, j \)\\
	RZ \( \gamma ,j \) \\
	CNOT \( i, j \)
\end{center}

The RZ is a single-qubit rotation gate; its parameter $\gamma$ is read from the classical track.
\subsubsection{Classical Control Logic Transformation}
The classical comparator in step 6 that determines whether \( E(m) < E_{\text{best}} \) is transformed into a Toffoli gate network via homomorphic mapping. Let \( a = E(m) \) and \( b = E_{\text{best}} \) be 32-bit fixed-point numbers; the comparator outputs \( c=1 \) if \( a < b \) and 0 otherwise. This Boolean circuit can be expressed as
\[
c = \bigvee_{i=31}^{0} \bigl( \neg a_i \land b_i \land \bigwedge_{j=i+1}^{31} (a_j = b_j) \bigr).
\]
Each logic gate (AND, OR, NOT) corresponds to a quantum gate (Toffoli, CNOT, X), and the entire circuit is reversible. When the classical head executes this circuit, each logic operation corresponds to one state transition. However, since this is performed on the classical track, unitarity is not violated (operations on the classical track are inherently deterministic Boolean functions, which HTM allows as special transitions).
\subsection{Resource Scheduling and Coordination Mechanisms}
The unified controller $C_f$ manages the concurrency and synchronization of the two read/write heads:
\begin{enumerate}
	\item Scheduling Strategy: When the quantum head executes long sequences of unitary transformations (steps 2-4), the classical head is in a H (halt) state and does not access the tape; after measurement is complete, the quantum head issues an interrupt, and the controller activates the classical head to execute the update steps.
	\item Resource Sharing: The classical track and the quantum track are isolated on the physical tape via address partitioning. The controller maintains an access arbiter: if both heads request the same address simultaneously (theoretically not possible, as the quantum head accesses only the quantum address segment and the classical head accesses only the classical address segment), the classical head is given priority (a deterministic strategy).
	\item Adaptive Iteration: After the classical\_circuit in step 6 calculates new parameters ($\gamma$',$\beta$'), the controller uses L (left shift) and R (right shift) instructions to position the tape head over the parameter storage area and overwrite the old values. Based on the value of k, it then decides whether to jump to  $q_{param}$  or $q_{term}$. This loop control is entirely realized by the finite state automaton of HTM, requiring no external intervention.
\end{enumerate}
\subsection{Complexity Analysis}
\subsubsection{Time Complexity}
For a single iteration, 
Quantum Part: \( U_{c}(\gamma) \) requires \( |E|=5 \) ZZ interactions, each compiled into two CNOT gates and one RZ gate, totaling fifteen gates; 
\( U_{M}(\beta) \) requires five RX gates; initial state preparation requires five H gates. Total gate count \( G=O(|E|+n)=O(10) \). In HTM, each gate corresponds to one step of the quantum head transition (assuming the quantum coprocessor executes one gate per step). Hence, the quantum time \( T_{q}=O(G) \).

Classical Part: Measuring and reading out 5 bits is constant time; calculating energy \( E(m) \) requires traversing 5 edges, each involving an XOR and addition, \( O(|E|) \). If analytical gradients (parameter shift rule) are used, a single quantum execution can simultaneously estimate the expectation value and its gradient. For simplicity, this analysis assumes 1 quantum execution per iteration plus \( O(|E|) \) classical computation.

Therefore, the total time for a single iteration is \( T_{\text{iter}}=G=O(|E|+n)=O(5) \), a constant time. With \( K=100 \) outer iterations, the total time is \( T_{\text{total}}=O(K \cdot (|E|+n))=O(500) \).
\subsubsection{Space Complexity}
Quantum Track: Storing a 5-qubit state requires 32 complex amplitudes. In a physical HTM implementation, this corresponds to 5 physical qubits (or a 32-dimensional vector in a simulator). The classical head does not copy the quantum state; it only accesses measurement results.

Classical Track: Stores parameters, counters, best energy value, etc., all \( O(1) \) classical words.

\subsubsection{Comparison with Classical Simulation}
If the same QAOA algorithm were simulated on a classical Turing machine, a 32-dimensional state vector would need to be explicitly maintained, and each unitary operation would require \( O(2^{2n}) = O(1024) \) complex arithmetic operations (matrix-vector multiplication). By using the quantum head to directly manipulate the physical quantum state, HTM reduces the computational complexity of the quantum part from exponential to polynomial (gate count linear in \( n \) and \( |E| \)). The classical control part remains polynomial in both cases. Therefore, HTM realizes the potential for exponential acceleration through classical-quantum collaboration.
\section{Conclusion}
HTM effectively alleviates the critical bottleneck of insufficient theoretical support in existing hybrid computing systems. By standardizing the state set and transition functions, designing collaborative control mechanisms between classical and quantum models, and conducting complexity analysis, HTM addresses technical challenges such as hybrid architecture design and control strategy. It also supports task partitioning and resource scheduling for the model, thereby filling an important theoretical gap in the field of classical-quantum hybrid computing.

To further advance the standardization, and practical deployment of the HTM model, future research may be carried out in the following directions: First, fully leverage the standardization and compatibility advantages of the HTM model, and continuously optimize and upgrade the model architecture, control protocols, and interaction interfaces in conjunction with hardware devices. Second, explore adaptive task partitioning strategies and resource scheduling algorithms to enhance its adaptability to more computing environments based more detailed complexity analysis.

\section*{Declarations}

\subsection*{Funding}
This work was supported by the National Key R\&D Program of China under Grant No.~2024YFB4504100, with the project led by Zheng~Shan.

\subsection*{Human Ethics Declaration}
Not applicable.
\bibliography{htmckwx0701}
\end{document}